\documentclass[openacc]{rsproca_new}

\begin{document}

\title{Viability of slow-roll inflation in light of the non-zero\\
$k_{\rm min}$ measured in the CMB power spectrum}

\author{Jingwei Liu$^1$ and Fulvio Melia$^{2}$}

\address{$^{1}$Department of Physics, The University of Arizona, AZ 85721, USA\\
$^2$Department of Physics, The Applied Math Program, and Department of Astronomy,
The University of Arizona, AZ 85721, USA}

\subject{Cosmology, Field Theory, Astrophysics}


\corres{F. Melia\\
\email{fmelia@email.arizona.edu}}

\begin{abstract}
Slow-roll inflation may simultaneously solve the horizon problem and
generate a near scale-free fluctuation spectrum $P(k)$. These two processes are
intimately connected via the initiation and duration of the inflationary phase.
But a recent study based on the latest {\it Planck} release suggests that
$P(k)$ has a hard cutoff, $k_{\rm min}\not=0$, inconsistent with this conventional
picture. Here we demonstrate quantitatively that most---perhaps all---slow-roll
inflationary models fail to accommodate this minimum cutoff. We show that the
small parameter $\epsilon$ must be $\gtrsim 0.9$ throughout the inflationary
period to comply with the data, seriously violating the slow-roll approximation.
Models with such an $\epsilon$ predict extremely red spectral indices, at odds
with the measured value. We also consider extensions to the basic
picture (suggested by several earlier workers) by adding a kinetic-dominated or
radiation-dominated phase preceding the slow-roll expansion. Our approach differs
from previously published treatments principally because we require these modifications
to---not only fit the measured fluctuation spectrum, but to simultaneously also---fix
the horizon problem. We show, however, that even such measures preclude a joint
resolution of the horizon problem and the missing correlations at large angles.
\end{abstract}


\begin{fmtext}
\section{Introduction}
The lack of large-angle correlation in the cosmic microwave background
(CMB) anisotropies, confir\-med by three independent satellite
missions \cite{Wright:1996,Bennett:2003,Planck:2018}, raises 
\end{fmtext}

\maketitle

\noindent serious questions concerning the viability of basic slow-roll inflation
\cite{Guth:1981,Linde:1982}. A reliance on cosmic variance \cite{Copi:2009}
for the missing correlations cannot avoid the correspondingly small
probabilities ($\lesssim 0.24\%$) that disfavor the conventional picture
at $\gtrsim 3\sigma$. This growing tension between the theoretical
predictions and the CMB observations was recently put on a much
more rigorous, formal footing with a detailed analysis of the
recent {\sl Planck} data release \cite{Planck:2018}, showing quite
robustly that the absence of large-angle correlation in the CMB
is due to a non-zero minimum wavenumber, $k_{\rm min}$, in the
fluctuation power spectrum $P(k)$ \cite{MeliaLopez:2018}.

The inflationary paradigm posits that quantum fluctuations were generated shortly
after the Big Bang \cite{Mukhanov:1992} with a power-law power spectrum $P(k)$
distributed over an indeterminate range of wavenumbers $k$. But the latest {\it Planck}
measurements are precise enough for us to question whether or not $k_{\rm min}$
is in fact zero. Ref.~\cite{MeliaLopez:2018} demonstrated that the lack of
large-angle correlation in the CMB is due to a cutoff $k_{\rm min}\not=0$,
and measured its value by optimizing the theoretical fits to the measured
angular-correlation function. These authors provided compelling evidence
that the {\sl Planck} data clearly rule out a zero $k_{\rm min}$ at
a very high level of confidence---{\sl exceeding} $8\sigma$. This measurement
is critically important because---given an inflaton potential, $V(\phi)$,
and the notion that a minimum wavenumber corresponds to the first mode
leaving the horizon---$k_{\rm min}$ signals a precise cosmic time,
$t_{\rm start}$, at which slow-roll inflation is supposed to have started.

Unconstrained slow-roll inflation would have stretched all fluctuations
beyond the horizon, resulting in a $P(k)$ with $k_{\rm min}=0$, which
would have produced strong correlations in the CMB at all angles, $\theta$,
in contrast to what is actually seen, i.e., an angular correlation function
that essentially goes to zero at $\theta\gtrsim 60^\circ$. The {\sl measured}
minimum wavenumber is instead
\begin{equation}
k_{\rm min}={4.34\pm0.50\over r_{\rm dec}}\;,
\end{equation}
where $r_{\rm dec}$ is the comoving distance between us and redshift
$z_{\rm dec}=1080$, at which decoupling in standard $\Lambda$CDM
cosmology is thought to have occurred. Therefore, for the latest
{\it Planck} parameters (see below), one finds $r_{\rm dec}\approx 13,804$~Mpc,
and a corresponding minimum wavenumber
\begin{equation}
k_{\rm min}=(3.14\pm 0.36)\times 10^{-4}\;{\rm Mpc}^{-1}\;.
\end{equation}

In the conventional inflationary picture, mode $k$ exited the
horizon at time $t_*$, satisfying the simple condition \cite{Mukhanov:1992}
\begin{equation}
{\lambda_k(t_*)\over 2\pi}={c\over H_*}\;,
\end{equation}
where $\lambda_k(t_*)={2\pi a(t_*)/k}$ is its wavelength, $a(t_*)$ is
the expansion factor in the Friedmann-Lema\^itre-Robertson-Walker metric (FLRW),
and $H_*$ is the Hubble constant at that moment. This strong observational
constraint therefore implies that standard slow-roll inflation must satisfy
the initial condition
\begin{equation}
a(t_{\rm start})H_{\rm start}=94.3\pm 10.9\;{\rm km}\;{\rm s}^{-1}\;{\rm Mpc}^{-1}\;.\label{con:cutoffconstraint}
\end{equation}
But as we shall show in this paper, at least some inflationary
models fail to solve the horizon problem in light of this new measurement.
We shall first consider pure slow-roll inflation on its own,
but then also demonstrate that the introduction of a kinetic-dominated (KD) or
radiation-dominated (RD) phase preceeding the slow-roll expansion cannot
produce consistency with the data either.

The missing angular correlation at large angles is related to
the unexpectedly low power measured in the small $\ell$ multiple moments.
Several workers have previously attempted to resolve this issue by
introducing such an RD or KD phase preceding the flatenning of the inflaton
potential. We shall summarize several of these efforts in \S~3 below, and
provide a set of pertinent references to this previously published work.
Our analysis in this {\it Letter} differs from many of these treatments
principally because we require such modifications to---not only account
for the missing angular correlation at large angles, but simultaneously
to also---fix the horizon problem. This caveat is critical to our
conclusion: that the measurement of $k_{\rm min}$ impacts both the
measured fluctuation spectrum and the ability of standard slow-roll
inflation to equilibrate the CMB temperatue across the visible Universe.

\section{Pure Slow-Roll Inflation}
We may clearly see the impact of this measurement by considering the simplest
case of a pure exponential (i.e., de Sitter) expansion. To ensure that the CMB
temperature seen today is equilibrated across the sky, a photon must
have traversed a comoving distance prior to decoupling at least twice
$r_{\rm dec}$. That is, the minimal condition for inflation is
\begin{equation}
r_{\rm preCMB}=2r_{\rm dec}\equiv 2c\int_{t_{\rm dec}}^{t_0}{dt\over a(t)}\;,
\end{equation}
where $a(t)$ is the aforementioned expansion factor. In terms of
$H=\dot{a}/a$, we may also put
\begin{equation}
r_{\rm dec}=c\int_{a_{\rm dec}}^{a_0}\frac{da}{a^2H}\;,
\end{equation}
where $H(a)$ is the Hubble parameter as a function of $a$, and $a_0$ is
the expansion factor today. The latest cosmological measurements all seem
to be consistent with a spatially flat Universe \cite{Planck:2018}, for
which $a_0$ may be normalized to $1$.

From the Friedmann equation, we have
\begin{equation}
H^2(a)=H_0^2\left(\frac{\Omega _{\rm m}}{a^3}+\frac{\Omega _{\rm r}}{a^4}+
\Omega_\Lambda\right)\;.
\end{equation}
Thus, for the {\it Planck} optimized values $H_0=66.99\pm0.92$ km s$^{-1}$ Mpc$^{-1}$,
$\Omega_{\rm m}=0.321\pm0.013$, $\Omega_\Lambda=0.679\pm0.013$, and
$\Omega_{\rm r}=9.3\times10^{-5}$ \cite{Planck:2018}, for the Hubble parameter,
and fractional matter and cosmological constant energy densities, respectively,
one finds $r_{\rm dec}\approx 13,804$~Mpc. By comparison, $r_{\rm preCMB}$ is
calculated from the start of inflation, $a_{\rm start}\equiv a(t_{\rm start})$,
to decoupling and is mostly due to the expansion up to $a_{\rm end}$,
when the inflaton field becomes sub-dominant. Thus,
\begin{equation}
r_{\rm preCMB}\approx c\int_{a_{\rm start}}^{a_{\rm end}}\frac{da}{a^2H}\;.
\end{equation}
In simple exponential (i.e., pure de Sitter) expansion, $H(a)=H_{\rm start}$
is constant during inflation, so
\begin{equation}
r_{\rm preCMB}=\frac{c}{H_{\rm start}}\left(\frac{1}{a_{\rm start}}-
\frac{1}{a_{\rm end}}\right)\;,
\end{equation}
and since $a_{\rm start}\ll a_{\rm end}$, we may also put
\begin{equation}
r_{\rm preCMB}=\frac{c}{H_{\rm start}a_{\rm start}}\;.
\end{equation}
The newly measured constraint in Equation~(1.4) therefore implies that
$r_{\rm preCMB}\approx 3,181$~Mpc, much smaller than the required comoving
distance $2r_{\rm dec}\approx 27,608$~Mpc. This factor $9$ disparity
therefore rules out pure exponential inflationary models, because they
could not solve the horizon problem given the measured value of $k_{\rm min}$.

But the focus today is on {\it slow-roll} inflation, for which $H(a)$ due
to the inflaton field is very nearly---though not exactly---constant. It
is not difficult to see that when the small parameter $\epsilon$ (see
Eq.~2.9 below) is monotonic \cite{Liddle:1994}, $H(a)\le H_{\rm start}$
for all $a\ge a_{\rm start}$. As such, one should expect
$r_{\rm preCMB}$ to be bigger than that in Equation~(2.6) (corresponding
to pure exponential expansion) if the starting condition (Eq.~1.4) remains
the same.

\begin{figure}[t]
\centering\includegraphics[width=4.0in]{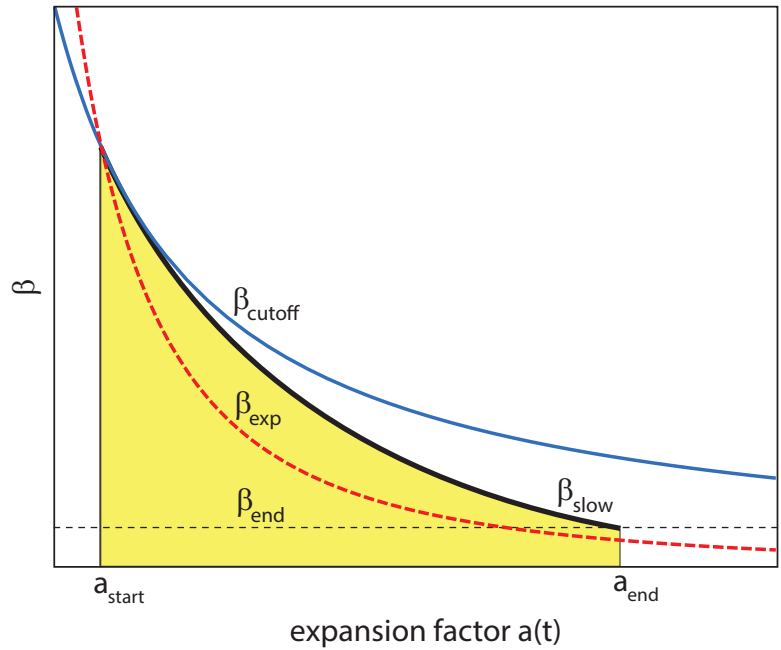}
\caption{Phase space of permitted $\beta(a)$ versus $a$ trajectories for
slow-roll inflationary models. Here, $\beta_{\rm cutoff}=(H_{\rm start}
a_{\rm start})^{-1}\,a^{-1}\propto 1/a$ (blue solid); $\beta_{\rm exp}=
(H_{\rm start})^{-1}a^{-2}\propto 1/a^2$ (red dashed); and $\beta_{\rm end}=
(H_{\rm end}a_{\rm end}^2)^{-1}=$ constant (black dashed). The shaded (yellow)
area is the dominant contribution to the integral for $r_{\rm preCMB}$, for
a specific slow-roll model with $\beta_{\rm slow}(a)$, and should therefore
be compared with the comoving distance $r_{\rm dec}$ to decoupling.}
\label{fig:1}
\end{figure}

To quantify the difference, let us define a new variable
\begin{equation}
\beta(a)\equiv\frac{1}{Ha^2}
\end{equation}
(i.e., the integrand in Eq.~2.4).
The boundaries relevant to the run of $\beta(a)$ with $a$ are shown
schematically in figure~\ref{fig:1}. The measured cutoff $k_{\rm min}$
corresponds to the solid blue hyperbola, on which $\beta_{\rm cutoff}=(H_{\rm start}
a_{\rm start})^{-1}\,a^{-1}\propto 1/a$. Inflation must begin at $a_{\rm start}$
somewhere on this curve. For example, if $H$ is constant (red dashed curve),
inflation initiates at the point where the solid and dashed curves
intersect, after which $\beta_{\rm exp}\propto 1/a^2$. Also, the Universe is
believed to have been radiation dominated right after inflation ended, for which
\begin{equation}
H^2_{\rm end}=H_0^2\left(\frac{\Omega_{\rm r}}{a_{\rm end}^4}\right)\;.
\end{equation}
Again, for exponential inflation with $H(a)=H_{\rm start}=H_{\rm end}$,
Equation~(2.8) corresponds to the horizontal (black) short-dash line, with
$\beta_{\rm end}=(H_{\rm end} a_{\rm end}^2)^{-1}=$ constant near the
bottom of the plot.
Any slow-roll inflationary model (with $H$ not exactly constant) would then
follow a trajectory $\beta_{\rm slow}(a)$ (shown as solid black) somewhere
between the $\beta_{\rm cutoff}$ and $\beta_{\rm exp}$ curves. It could never
cross the hyperbola because $H$ can never be bigger than its starting value
$H_{\rm start}$.

The small parameter $\epsilon$ is defined according to \cite{Liddle:1994}
\begin{equation}
\epsilon\equiv\frac{m_{\rm Pl}^2}{4\pi}\left(\frac{H'}{H}\right)^2\;,
\end{equation}
where $m_{\rm Pl}$ is the Planck mass and prime denotes a derivative
with respect to the inflaton scalar field, $\phi$. It is not difficult
to show that
\begin{equation}
H(\phi)=H_{\rm start}\exp\left(-\int_{\phi_{\rm start}}^{\phi}
\sqrt{\frac{4\pi\epsilon(\phi)}{m_{\rm Pl}^2}}\, d\phi\right)\;,
\end{equation}
where the subscript `start' has its usual meaning. It is also useful
to introduce the number of e-folds during inflation,
\begin{equation}
N(\phi_{\rm start},\phi)\equiv\ln\left(\frac{a}{a_{\rm start}}\right)=
\int_{\phi_{\rm start}}^{\phi}\sqrt{\frac{4\pi}{m_{\rm Pl}^2\,\epsilon(\phi)}}\, d\phi\;.
\end{equation}
Clearly, $\epsilon=0$ if $H$ is strictly constant. It is non-zero, but
small, if $H$ changes slowly (hence the designation `slow-roll'). Thus,
inflation in slow-roll models must end when $\epsilon$ increases to $1$,
at which point the slow-roll approximation breaks down.

Let us therefore first consider the extreme case in which $\epsilon=1$
throughout the inflationary phase, for which
\begin{equation}
H(a)=H_{\rm start}\exp(-N)=\frac{H_{\rm start}a_{\rm start}}{a}\;.
\end{equation}
This is in fact the solid black hyperbola shown in figure~\ref{fig:1}.
Therefore,
\begin{equation}
r_{\rm preCMB}^{\epsilon=1}=\frac{c}{H_{\rm start}a_{\rm start}}
\ln\left(\frac{a_{\rm end}}{a_{\rm start}}\right)\;.
\end{equation}
This comoving distance is bigger by a factor $\ln(a_{\rm end}/a_{\rm start})$
than that for pure de Sitter expansion (Eq.~2.6), and would be
sufficient to account for the required value of $2r_{\rm rec}$. As we shall
discuss shortly, however, there are compelling reasons why such a
persistently large value of $\epsilon$ is inconsistent with the data.
Typically, slow-roll models have a very tiny $\epsilon$ during most of
inflation, approaching 1 only towards the end, when the inflaton field
is believed to somehow dissolve into standard model particles, so that
the magnitude of $H^\prime$ becomes very large.

To more accurately represent such models, we therefore define another new parameter
$0<b<1$ such that $\epsilon^2$ is restricted to values $\le b$ during most of
the inflationary expansion, breaking down only at the very end. Then we have
\begin{eqnarray}
H(\phi)&>&H_{\rm start}\exp\left(-\int_{\phi_{\rm start}}^{\phi}
\sqrt{\frac{4\pi b}{m_{\rm Pl}^2\epsilon(\phi)}}\,d\phi\right)\nonumber \\
&=&H_{\rm start}\exp(-\sqrt{b}N)\nonumber \\
&=&H_{\rm start}\left(\frac{a}{a_{\rm start}}\right)^{-\sqrt{b}}\;,
\end{eqnarray}
so that, assuming $a_{\rm end}>>a_{\rm start}$,
\begin{equation}
r_{\rm preCMB}^{\epsilon^2<b}<r_{\rm preCMB}^{\epsilon^2=b}
\equiv\frac{1}{(1-\sqrt{b})H_{\rm start}
a_{\rm start}}\label{con:epsilonandc}
\end{equation}
and, combining this with Equation~(2.2),
we find that $\sqrt{b}>0.875$ in order for the right-hand side of
Equation~(2.15) to exceed $2r_{\rm dec}$ and solve the horizon problem.

In other words, $\epsilon$ must be quite large compared to typical
values required in commonly studied slow-roll models. Indeed, scenarios with
$\epsilon\sim 1$ during the whole of inflation have already been considered
and eliminated on observational grounds \cite{Cook:2016}, because either
(i) inflation would not have lasted long enough to fix the horizon problem,
or (ii) the predicted extremely red spectral index ($n_s\ll 1$) in $P(k)$
would be substantially different from its observed value $0.9649\pm 0.0042$
\cite{Planck:2018}. Inflationary models with $\epsilon^2>b$ are therefore
not at all practical.

\begin{figure}[!h]
\centering\includegraphics[width=4.0in]{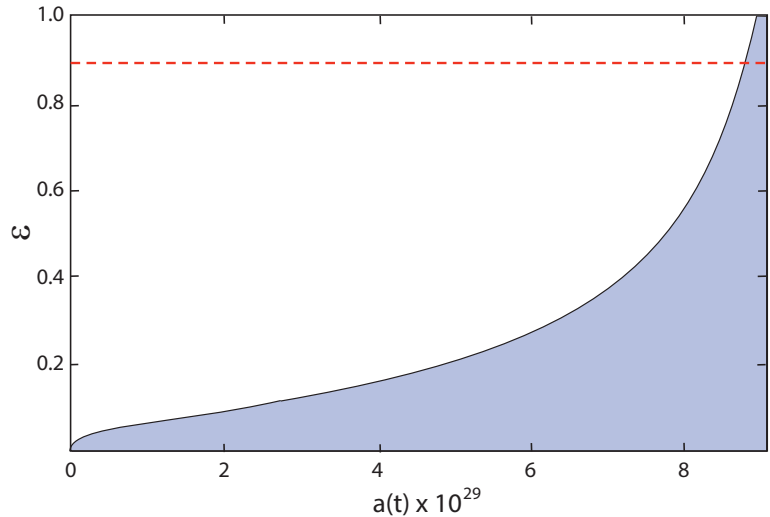}
\caption{The small parameter $\epsilon$ as a function of the expansion
factor $a(t)$ for the `small-field' inflaton potential in Equation~(2.16).
The (red) dashed line marks the value required for the model to comply
with the {\it Planck} measurement of $k_{\rm min}$.}
\label{fig:2}
\end{figure}

To demonstrate this general result more practically, let us 
examine its impact on four rather well-known, specific types of potential that 
have been studied thus far, beginning with the evolution of the slow-roll 
parameter $\epsilon$ in so-called `small-field' inflation models, for which 
the potential may be approximated locally by the expression
\begin{equation}
V(\phi)=V_0\left[1-\left(\phi/\mu\right)^p\right]\;.
\end{equation}
As an illustration, we take $p=2$ and $\epsilon(a_{\rm end})=1$ (the value of 
$\mu$ is irrelevant for the calculation of $\epsilon[a]$). Higher-order terms
in $V(\phi)$ become important only towards the end of inflation. Our numerical 
solution for $\epsilon$, based on the {\it Planck} optimized parameter values
(see paragraph following Eq.~2.3 above), is shown in figure~\ref{fig:2}. Indeed, 
$\epsilon^2>b$ for $8.5\times 10^{-29}\lesssim a\lesssim a_{\rm end}= 
8.9\times 10^{-29}$, but is far too small elsewhere for $r_{\rm preCMB}$ 
to exceed $2r_{\rm rec}$.

An alernative characterization of such an evolution may be written
in terms of the number of e-folds (Eq.~2.11) required during inflation in order
to overcome the horizon problem, compared to the actual number permitted by the
$k_{\rm min}$ constraint in Equation~(1.4). We have numerically calculated
$\epsilon_V$ (see Eq.~2.19 below) and $N$, subject to this constraint, for the 
following three slow-roll potentials:
\begin{eqnarray}
V_Q(\phi)&=&\frac{1}{2}m\phi^2\qquad\qquad\quad\;\;\;{\rm (Quadratic)}\nonumber\\
V_H(\phi)&=&V_0[1-(\frac{\phi}{\mu})^2]^2\qquad\;{\rm (Higgs-like)}\nonumber\\
V_N(\phi)&=&V_0[cos(\frac{\phi}{f})+1]\qquad{\rm (Natural)}\;.
\end{eqnarray} 
For specificity, we have continued to use the {\it Planck} optimized parameters. The 
principal difference between our calculation and those carried out in previous work 
is the inclusion of Equation~(1.4) as an initial condition. In addition, to this
constraint, the other inputs informing the calculation include: (1) the observed
value of the scalar spectral index, $n_s=0.96$, which was measured at the pivot point
$k_{\rm pivot}\equiv 0.05$ Mpc$^{-1}$ \cite{Planck:2018}; (2) an endpoint of inflation
at $\epsilon_V=1$ (see Eq.~2.19 below); and (3) a smooth transition from this inflated
expansion to one driven by a radiation-dominated equation-of-state, as shown in 
Equation~(2.8). 

\begin{figure}[t]
\centering\includegraphics[width=4.0in]{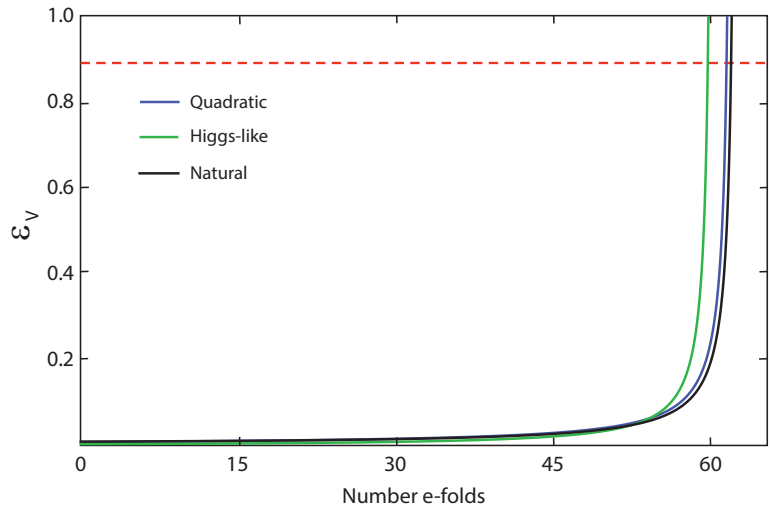}
\caption{The small parameter $\epsilon_V$ as a function of the number
of e-folds (Eq.~2.11) for three illustrative slow-roll potentials given in Equation~(2.17).
The (red) dashed line marks the value required for the model to comply with the 
{\it Planck} measurement of $k_{\rm min}$.}
\label{fig:3}
\end{figure}

At the early stage of inflation, these potentials may be used to define an alternative
set of `small parameters' $\eta_V$ and $\epsilon_V$ \cite{Liddle:1994}, such that
\begin{equation}
n_s\approx 2\eta_V-6\epsilon_V\;,
\end{equation}
where
\begin{eqnarray}
\epsilon_V&=&{m_{\rm Pl}^2\over 16\pi}\left({V^\prime\over V}\right)^2\nonumber\\
\eta_V&=&{m_{\rm Pl}^2\over 8\pi}{V^{\prime\prime}\over V}\;.
\end{eqnarray}
The approximation breaks down when $\epsilon_V$ is large, which is conventionally taken
to indicate the end of the inflated expansion. The key results of our simulations are as 
follows:

\begin{itemize}
\item {\it Quadratic:} $r_{\rm preCMB}=5,547$ Mpc, which is still a factor $\sim 5$ too 
small compared to $2r_{\rm dec}=27,608$ Mpc. This potential would have expanded the Universe 
by 62 e-folds (see fig.~\ref{fig:3}), but 64 e-folds would have been required to fix the horizon
problem. The difference of 2 e-folds accounts for the factor 5 difference between
$r_{\rm preCMB}$ and $2r_{\rm dec}$.

\item {\it Higgs-like:} $r_{\rm preCMB}=3,339$ Mpc, which is a factor $\sim 8$ too small.
In this case, the Universe would have expanded by 60 e-folds (fig.~\ref{fig:3}), but a little over
62 e-folds would have been required to mitigate the horizon problem.

\item {\it Natural:} $r_{\rm preCMB}=3,650$ Mpc, which is also a factor $\sim 8$ too small.
The Universe would have expanded by 63 e-folds, but a little over 65 e-folds would have been 
required to completely mitigate the horizon problem.
\end{itemize}

For direct comparison, the small parameter $\epsilon_V$ is shown as a function of $N$ for each of 
these three inflaton potentials in figure~\ref{fig:3}. As discussed earlier, inflation would have
ended when $\epsilon_V\rightarrow 1$. As was the case in figure~\ref{fig:2}, the horizontal red (dashed)
line indicates the approximate value $\epsilon_V$ requires to comply with the {\it Planck} measurement
of $k_{\rm min}$, and we see that, while $\epsilon_V$ does cross this mark in each case, it
is not sustained at this high level long enough for the Universe to have expanded sufficiently
to remove the horizon problem.

\section{Slow-roll Inflation Preceeded by a KD or RD Fast-roll Phase}
It appears, therefore, that to simultaneously resolve both
the horizon problem and the missing correlations at large angles, one must
consider additional inflationary phases coupled to the standard slow-roll expansion.
A closely related problem to the missing correlations at large angles is the
observed lack of power on the largest scales. Several authors have previously
attempted to mitigate this problem by introducing additional features to inflation,
such as the aforementioned KD and RD phases. For example, ref.~\cite{Destri:2008}
showed that an early fast-roll inflation can lead to a depression of the cosmic microwave
background quadrupole moment, with a characteristic scale $k_1\sim (3,759\;{\rm Mpc})^{-1}$
of the implied attractive potential. This is consistent with our previously measured minimum
cutoff $k_{\rm min}= (3,442\;{\rm Mpc})^{-1}$. These authors did not, however,
simultaneously calculate the comoving distances $r_{\rm preCMB}$ and $r_{\rm dec}$
to ensure that $r_{\rm preCMB}\ge 2r_{\rm dec}$. Subsequent work by these authors
\cite{Destri:2010} to include both a decelerated fast-roll and an inflationary fast-roll
phase similarly did not address the horizon problem in terms of the required comoving
distances. In addition, this work appears to rely on the Bunch-Davies initial conditions,
which may be problematic in the context of trans-Planckian physics.

This general approach was followed by other authors \cite{Scacco:2015}, who found
that a fast-rolling KD initial phase improves the primordial power spectral fit to the
data, but they similarly did not consider the impact of this treatment on $r_{\rm preCMB}$
versus $r_{\rm dec}$. Likewise, the Planck Collaboration \cite{Planck:2015} considered
the impact of a cutoff on the spectrum, though not the angular-correlation function.
Their treatment apparently also lacks a discussion of the possible impact of such a
cutoff on $r_{\rm preCMB}$ and $r_{\rm dec}$.

The work of ref.~\cite{Santos:2018} was published after our measurement of
$k_{\rm min}$ \cite{MeliaLopez:2018}, and they too considered the impact of a sharp
cutoff to the fluctuation spectrum. They concluded that the standard power-law
is preferred by the data, but made no mention of the horizon problem and
the lack of correlations at large angles, however, and the impact of this approach on
$r_{\rm preCMB}$ versus $r_{\rm dec}$.

An early phase of KD inflation was also introduced in
ref.~\cite{Handley:2014}, though restricted to only polynomial and exponential
potentials. These authors confirmed that such a transition exhibits a generic
damping of power on large scales, but did not explicitly consider its impact
on the angular correlation function and $r_{\rm preCMB}$ versus $r_{\rm dec}$.

The work that comes closest in spirit to our analysis in this paper is that reported
in ref.~\cite{Ramirez:2012a}. These authors, however, considered specifically
the $\lambda\phi^4$ potential and imposed the condition of ``just-enough"
inflation. They found that the slow-roll conditions are violated at the largest
scales, and that this approach cannot explain the lack of power at
the largest angles. In subsequent work \cite{Ramirez:2012b}, this treatment
was expanded to include quadratic and hybrid-type potentials, but still without
a consideration of their impact on the angular correlation function.

Finally, ref.~\cite{Remmen:2014} analyzed how much inflation one should expect
for a given energy scale of order $10^{16}$ GeV. But this work lacks direct
relevance to our proposed coupling of $k_{\rm min}$ measured from the angular
correlation function to the number of e-folds itself, and its bearing on
$r_{\rm preCMB}$ versus $r_{\rm dec}$.

Quite clearly, many authors have by now noted the glaring inconsistency associated
with low power in the CMB fluctuations on large scales, which is closely related to their lack
of correlation at large angles. Our work amplifies this general view by providing a much
stronger argument for a cutoff $k_{\rm min}$ in the primoridal fluctuation spectrum,
and its direct impact also on the horizon problem itself. To complete this discussion,
we shall now consider whether a KD or RD modification to the basic slow-roll inflationary
picture can help mitigate the inconsistency between $r_{\rm preCMB}$ and $r_{\rm dec}$
when a cutoff $k_{\rm min}$ is invoked to suppress the correlation
at large angles.

We shall first follow a simplified approach in which we gauge the impact
of a KD or RD modification to the horizon problem based solely on the previously
measured hard cutoff $k_{\rm min}$. It is well known, however, that the angular
power $C_\ell$ of each multipole $\ell$, from which the angular correlation function
$C(\theta)$ is calculated, depends on the entire fluctuation spectrum $P(k)$
\cite{MeliaLopez:2018}. Thus, any modification to the power spectrum produced
during the KD or RD phase alters $C(\theta)$ from that expected under pure
slow-roll conditions. Following our initial discussion of the impact of KD or RD
on the horizon problem using the previously measured $k_{\rm min}$, we shall
therefore quantitatively assess how much the cutoff wavenumber changes when
the angular correlation function is re-optimized for a representative inflaton
potential that contains a KD phase transitioning into slow-roll at $k_{\rm start}$.
We shall find that $k_{\rm start}$, signalling the start of inflated expansion,
can differ fractionally from $k_{\rm min}$ when $C(\theta)$ is fit to the {\it Planck}
data, though insufficiently to qualitatively alter any of the results.

We begin with a radiation-dominated Universe from the Big Bang to the onset
of inflation, during which
\begin{equation}
H=Qa^{-2}\;,
\end{equation}
where $Q$ is a constant. Solving for the scale factor, one therefore has
\begin{equation}
a^2=2Qt\;,
\end{equation}
so that
\begin{equation}
dt=\frac{a\,da}{Q}\;.
\end{equation}
The comoving distance traveled by a photon during this period is therefore
\begin{equation}
r_{\rm RD}=c\int_{0}^{t_{\rm start}}\frac{dt}{a}=\frac{ca_{\rm start}}{Q}\;.
\end{equation}
Thus, combining this with Equation~(3.1), we have
\begin{equation}
r_{\rm RD}=\frac{c}{H_{\rm start}\,a_{\rm start}}\;.
\end{equation}
The addition of an RD period preceeding slow-roll inflation can therefore double the comoving distance travelled by
a photon prior to the end of the inflation. Even this, however, is still far too small to solve the horizon problem, which
requires the comoving  distance to be at least 10 times bigger.

The addition of a KD fast-roll expansion may hold more promise. For such a scalar field-dominated Universe, we
have \cite{Liddle:1994}:
\begin{equation}
H(\phi)^2=\frac{8\pi}{3m_{\rm Pl}^2}\left(\frac{1}{2}\dot\phi^2+V(\phi)\right)\;,
\end{equation}
and
\begin{equation}
\ddot\phi+3H\dot\phi+V^\prime=0\;.
\end{equation}
From these two expressions, we derive
\begin{equation}
\dot H=-\frac{4\pi}{m_{\rm Pl}^2}\dot\phi^2\;,
\end{equation}
and
\begin{equation}
\dot\phi=-\frac{m_{\rm Pl}^2}{4\pi}H^\prime\;.
\end{equation}
For a KD scalar-field potential, Equation~(3.6) reduces to
\begin{equation}
H(\phi)^2\approx\frac{8\pi}{6m_{\rm Pl}^2}\dot\phi^2
\end{equation}
and, solving for $H$, we find that
\begin{equation}
H(\phi)=H_{\rm start}e^{\frac{2\sqrt{3\pi}}{m_{\rm Pl}}(\phi-\phi_{\rm start})}\;,
\end{equation}
for which
\begin{equation}
H^\prime=\frac{2\sqrt{3\pi}}{m_{\rm Pl}}H_{\rm start}e^{\frac{2\sqrt{3\pi}}{m_{\rm Pl}}(\phi-\phi_{\rm start})}\;.
\end{equation}

With Equation~(3.9), we therefore find that
\begin{equation}
\frac{dt}{d\phi}=-\sqrt{\frac{\pi}{3}}\frac{2}{m_{\rm Pl}H_{\rm start}}e^{\frac{2\sqrt{3\pi}}{m_{\rm Pl}}(\phi_{\rm start}-\phi)}\;,
\end{equation}
so that
\begin{equation}
t-t_i=\frac{1}{3H_{\rm start}}e^{\frac{2\sqrt{3\pi}}{m_{\rm Pl}}(\phi_{\rm start}-\phi)}\;,
\end{equation}
where $t_i$ is the time at which the KD expansion begins.
Thus, with
\begin{equation}
\tilde t\equiv t-t_i\;,
\end{equation}
we also have
\begin{equation}
\tilde t=\frac{1}{3H}\;,
\end{equation}
during this phase prior to the onset of slow-roll inflation.

We may now solve for the scale factor $a(t)$, finding that
\begin{equation}
a=M{\tilde t}^{\,1/3}\;,
\end{equation}
so that
\begin{equation}
dt=d\tilde t=\frac{3a^2}{M^3}\,da\;,
\end{equation}
where $M$ is another constant.
Therefore, the comoving distance travelled by a photon during this period is
\begin{equation}
r_{\rm KD}=c\int_{{\tilde t}_i}^{\tilde t_{\rm start}}\frac{d\tilde t}{a}
=\frac{3c}{2M^3}(a_{\rm start}^2-a_i^2)\;.
\end{equation}

As long as the KD period begins right after the Big Bang, we may therefore approximate this expression as
\begin{equation}
r_{\rm KD}\approx\frac{3c}{2M^3}a_{\rm start}^2\;,
\end{equation}
and therefore we find, with the use of Equations~(3.11) and (3.12), that
\begin{equation}
r_{\rm KD}\approx\frac{c}{2a_{\rm start}H_{\rm start}}\;.
\end{equation}
Clearly, even combining this comoving distance with that from the slow-roll inflationary
period, we find that $r_{\rm preCMB}$ is still far too small to solve the horizon problem.

Finally, we consider all three phases together, beginning with an RD period, followed by a KD
Universe and a subsequent slow-roll expansion. It is not difficult to show that
\begin{equation}
r_{\rm RD+KD}=\frac{3c}{2M^3}\left(a_{\rm start}^2-a_\star^2\right)+\frac{ca_{\rm \star}}{Q}\;,
\end{equation}
where $a_\star$ is the scale factor at the RD to KD transition. Thus
\begin{equation}
r_{\rm RD+KD}=\left(1+\frac{a_{\rm \star}^2}{a_{\rm start}^2}\right)\left(\frac{c}{2a_{\rm start}H_{\rm start}}\right)
<\frac{c}{a_{\rm start}H_{\rm start}}\;.
\end{equation}

In the last step, we examine the possibility that a more careful calculation of the
angular correlation function $C(\theta)$ with a modified $P(k)$ from the KD phase may yield an
optimized wavenumber $k_{\rm start}$ (signalling the start of inflation) differing
from the hard cutoff $k_{\rm min}$ we have been using in this analysis. It is not difficult
to show from Equations~(3.10-3.12) that the fluctuation spectrum produced during KD is
$P(k)\sim k^3$. To estimate the change one should expect to see with this more detailed
approach, we therefore now proceed to re-optimize $C(\theta)$ with
\begin{equation}
    P(k) = \left\{\begin{array}{ll}
        A_s(k/k_0)^{n_s-1}  &  {\rm if} k\ge k_{\rm start} \\
        A_s(k_{\rm start}/k_0)^{n_s-4}(k/k_0)^3 & {\rm if} k< k_{\rm start}\;,
        \end{array}\right.
\end{equation}
invoking the usual pivot scale $k_0$.

We follow the procedure outlined in ref.~\cite{MeliaLopez:2018}, and infer that
the angular power of multipole $\ell$ relevant to the Sachs-Wolfe domain of fluctuations
may be approximated as
\begin{equation}
C_\ell=B\int_0^{u_{\rm start}}\left({u\over u_{\rm start}}\right)^3{j_\ell^2(u)\over u}\,du+
B\int_{u_{\rm start}}^\infty {j_\ell^2(u)\over u}\,du\;,
\end{equation}
where $B$ is a normalization constant encompassing $A_s$ and several other factors; the
variable $u$ is defined by the expression $u\equiv k r_{\rm dec}$, in terms of the comoving
distance $r_{\rm dec}$ to the decoupling surface; and $j_\ell$ is the spherical Bessel
function of order $\ell$. The angular correlation function itself is then given by the expression
\begin{equation}
C(\theta)=\sum_\ell{(2\ell+1)\over 4\pi}C_\ell\,P_\ell(\cos\theta)\;,
\end{equation}
where $P_\ell(\cos\theta)$ are the Legendre polynomials \cite{Bond:1984}.

\begin{figure}[t]
\centering\includegraphics[width=4.0in]{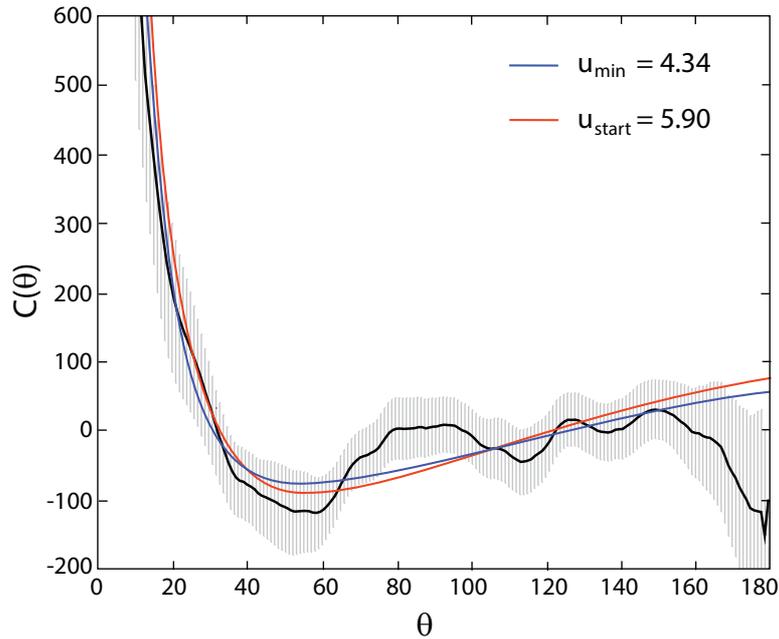}
\caption{The best-fit angular correlation functions for $P(k)$ with a hard cutoff
$k_{\rm min}$ (blue) and for $P(k)$ given in Equation~(3.24) (red), with an optimized
value $k_{\rm start}=4.12\times 10^{-4}\;{\rm Mpc}^{-1}$. These theoretical curves
are compared to the angular correlation function measured with {\it Planck} 
(dark solid curve) \cite{Planck:2015}, and associated $1\sigma$ errors (grey). (Adapted
from ref.~\cite{MeliaLopez:2018})}
\label{fig:4}
\end{figure}

Using Equation~(3.26) to refit the angular correlation function measured by {\it Planck}
\cite{Planck:2015,MeliaLopez:2018}, we find that the optimized fit corresponds to the value
$u_{\rm start}=5.9$. Thus, according to the definition of $u$, we find that
\begin{equation}
k_{\rm start}=4.12\times 10^{-4}\;{\rm Mpc}^{-1}\;.
\end{equation}
In figure~\ref{fig:4}, we show a comparison of the optimized angular correlation
functions for $P(k)$ with a hard cutoff $k_{\rm min}$ (blue) and $P(k)$
given in Equation~(3.24) (red) with this $k_{\rm start}$. The curves are almost
indistinguishable, though the blue one is a slightly better fit to the {\it Planck}
data at both small ($\theta\lesssim 45^\circ$) and large ($\theta\gtrsim 120^\circ$) 
angles \cite{MeliaLopez:2018}. This difference, however, is
too small for us to decide which of these fluctuation distributions is preferred
by the {\it Planck} data. Instead, the principal outcome of this comparison is the
change in wavenumber signalling the initiation of inflated expansion: from $k_{\rm min}$
in Equation~(1.2) for the hard cutoff, to $k_{\rm start}$ in Equation~(3.27) for the KD plus
slow-roll potential.

Thus, replacing $k_{\rm min}$ in Equation~(1.3) with $k_{\rm start}$, and using
Equations~(2.6) and (3.23), we find that $r_{\rm preCMB}<4,848$ Mpc, which is still
much smaller than the value (i.e., $27,608$ Mpc) required to solve the horizon
problem. In effect, the more detailed treatment of $P(k)$ has increased $r_{\rm preCMB}$
by about $50\%$, but nowhere near the factor $\sim 9$ required for this purpose.

No matter when the transition from RD to KD would have occurred, we find that no such
modification to the basic slow-roll scenario can render inflation consistent with the
measured $k_{\rm min}$ cutoff in the primordial fluctuation spectrum. The key point here is that,
while introducing a cutoff to the fluctuation distribution can account for the observed
CMB anisotropies, it cannot simultaneously solve the horizon problem.

\section{Conclusion}
The most recent {\it Planck} data have affirmed the absence of large-angle
correlations in the CMB anisotropies, seen previously with several instruments over several 
decades. A prevailing view is that this feature may simply be due to `cosmic variance,' 
based on the reasonable argument that we have only one Universe to observe, and that a variation
away from its most probable configuration should not be unexpected. Certainly none of the work
reported in this paper can completely eliminate that possibility. Nevertheless, seeking to find
alternative explanations, as we have attempted to do here, is motivated by the presumed low probability
of cosmic variance being the sole answer. The analysis reported in ref.~\cite{MeliaLopez:2018} shows
that a more probable explanation for the lack of large-angle correlations in the CMB is the presence 
of a hard cutoff $k_{\rm min}$ in the $P(k)$ spectrum. If true, this cutoff has profound consequences 
on the viability of slow-roll inflationary models because $k_{\rm min}$ points to a well-defined 
time at which inflation could have started. Quantifying this impact on the possible form of the
inflaton potential has been the main goal of this paper.

The constraint implied by $k_{\rm min}$ allows inflation to simultaneously solve the horizon 
problem and produce a near power-law fluctuation spectrum only if $\epsilon\approx 1$ throughout 
the inflationary expansion. But such a scenario then predicts an extremely red spectral index 
completely at odds with the measured value. Here, we have examined in detail the consequences
of $k_{\rm min}$ on four well-studied slow-roll inflationary models proposed thus far, 
showing that, if our interpretation of $k_{\rm min}$ is correct, the {\it Planck} CMB data 
rule out such slow-roll potentials at a very high level of confidence.

\section*{Acknowledgments}
FM is grateful to Amherst College for its support through a John
Woodruff Simpson Lectureship.

\vskip 0.2in\noindent{\bf Ethics Statement.} This research poses no ethical considerations.

\vskip 0.2in\noindent{\bf Data Accessibility Statement.} All data used in this paper were
previously published.

\vskip 0.2in\noindent{\bf Competing Interests Statement.} We have no competing interests.

\vskip 0.2in\noindent{\bf Authors' contributions.} The authors together conceived the project,
carried out the calculations, and wrote the paper. All authors gave final approval for
publication.

\vskip 0.2in\noindent{\bf Funding.} None.

\end{document}